\documentclass[prb, amsmath, amssymb, superscriptaddress, twocolumn]{revtex4-2}

\usepackage{graphicx,xcolor}

\usepackage[colorlinks,plainpages=false,linkcolor=red,urlcolor=blue,citecolor=blue,pdfpagemode=UseNone,pdfstartview=FitBH]{hyperref}

\newcommand{\m}{\mathbf}

\newcommand{\om}{\omega}

\newcommand{\be}{\begin{eqnarray}}
\newcommand{\ee}{\end{eqnarray}}
\newcommand{\nn}{\nonumber}

\newcommand{\gdru}{GdRu$_2$Si$_2$}

\begin{document}

\title{Thermodynamic theory of square skyrmion lattice in tetragonal frustrated antiferromagnets}

\author{Oleg I. Utesov}
\affiliation{Center for Theoretical Physics of Complex Systems, Institute for Basic Science (IBS), Daejeon 34126, Republic of Korea}
            
\author{Danila P. Budylev}
\affiliation{National Research University ``Higher School of Economics'', Moscow 109028, Russia}

\affiliation{The Faculty of Physics of St. Petersburg State University, St. Petersburg 198504, Russia}

\begin{abstract}

High-temperature part of the phase diagram of tetragonal frustrated antiferromagnets is discussed within the framework of the mean-field approach. Based on recent experimental findings, we generalize previous theoretical studies by considering a model that includes frustrated Heisenberg exchange, biquadratic exchange, magnetodipolar interaction,  anisotropic exchange, and single-ion anisotropy. It is analytically demonstrated that a subtle interplay among these interactions results in a variety of phase diagrams in the temperature-magnetic field plane. We argue that one of the proposed diagrams reproduces all crucial features of the phase diagram experimentally observed for~\gdru~compound. Besides magnetodipolar interaction and additional easy-axis contribution, it requires moderate biquadratic exchange. We show that despite the remarkable square skyrmion lattice being stable even if only magnetodipolar interaction/compass anisotropy or biquadratic exchange is included, their ``symbiosis'' allows for greatly enhancing its stability region. It is also demonstrated that higher-order harmonics play an important role in the stabilization of the square skyrmion lattice. The developed analytical approach can be useful for the refinement of the microscopic model parameters when comparing its predictions with experimental findings.

\end{abstract}

%\begin{keyword}

%Square skyrmion lattice \sep  Topological phases  \sep Phase %diagram \sep Mean-field approach \sep Biquadratic exchange

%\end{keyword}

%\end{frontmatter}

\maketitle

\section{Introduction}
\label{Sintro}

Skyrmions and their ordered arrays -- skyrmion lattices (SkL) -- are among the most studied objects of contemporary condensed matter physics~\cite{bogdanov2020}. Skyrmions were first predicted as metastable states of 2D ferromagnets~\cite{belavin1975metastable}. A possibility to stabilize isolated skyrmions due to Lifshitz invariants (Dzyaloshinskii-Moriya interaction~\cite{dzyaloshinsky1958,moriya1960}) was shown in Ref.~\cite{bogdanov1989}. Next, it was demonstrated that skyrmions can form an ordered state, a hexagonal skyrmion lattice, under certain conditions~\cite{bogdanov1994}. Notably, this superstructure was indeed observed in MnSi using neutron scattering~\cite{muhlbauer2009} and the Hall effect~\cite{neubauer2009}, which caused an upsurge of interest in topological spin structures. It is also stimulated by prospective applications, e.g., the racetrack memory~\cite{fert2013,fert2017} or unconventional computing~\cite{li2021magnetic}.

The efficiency of various applications relies on magnetic skyrmions' non-trivial topology~\cite{belavin1975metastable} and, hence, topological protection from local perturbations. The topological charge of a spin structure is defined using the spin winding number on a unit sphere:
\begin{equation}\label{charge1}
  Q = \frac{1}{4 \pi} \int \m{n} \cdot \left[ \partial_x \m{n} \times \partial_y \m{n} \right] dx dy,
\end{equation}
where $\m{n} = \m{s}/|\m{s}|$ is a unit vector along the averaged over thermodynamical and quantum fluctuations spin direction. For a single skyrmion, the integral usually yields $Q = \pm 1$, whereas for SkL the natural measure is a density of the topological charge, $n_\textrm{sk}$. The latter quantity is crucial. For instance, it defines the topological Hall resistivity ($\rho^T_{xy} \propto n_\textrm{sk}$)~\cite{neubauer2009}. Nowadays, other topologically nontrivial objects, e.g., meron-antimeron structures, are also actively studied~\cite{yu2018,gobel2021beyond,hayami2021meron,kim2023emergence,kuchkin2023heliknoton,leonov2024meron}.

Remarkably, frustration of exchange interaction can lead to the stabilization of short-period magnetic structures with a high density of topological charge $n_\textrm{sk}$~\cite{leonov2015}, leading to the giant topological Hall effect. Compact skyrmion lattices with periods of several nanometers were indeed observed in triangular Gd$_2$PdSi$_3$~\cite{kurumaji2019SkL}, breathing kagom\'{e} Gd$_3$Ru$_4$Al$_{12}$~\cite{hirschberger2019skyrmion}, and tetragonal \gdru~\cite{khanh2020}, GdRu$_2$Ge$_2$~\cite{yoshimochi2024multistep}, EuAl$_4$~\cite{takagi2022square}. Recent experimental and theoretical progress on this topic is reviewed in Ref.~\cite{kawamura2025frustration}.

Shortly after the square skyrmion lattice was observed in~\gdru~(see Ref.~\cite{khanh2020}), three theoretical studies addressing mechanisms of its stabilization were published~\cite{hayami2021square,utesov2021tetragonal,wang2021}. It was shown that important ingredients of the corresponding microscopic models are biquadratic exchange~\cite{takahashi1977half,spivsak1997theory,hayami2017} and complex anisotropic interactions. The latter include compass anisotropy, anisotropic exchange, and magnetodipolar interaction. The subsequent studies were devoted to the identification of various magnetic phases~\cite{khanh2022,wood2023}, electronic structure, and exchange couplings~\cite{bouaziz2022,wood2025magnon} of~\gdru.

\begin{figure*}[t]
    \centering
    \includegraphics[width=0.95\linewidth]{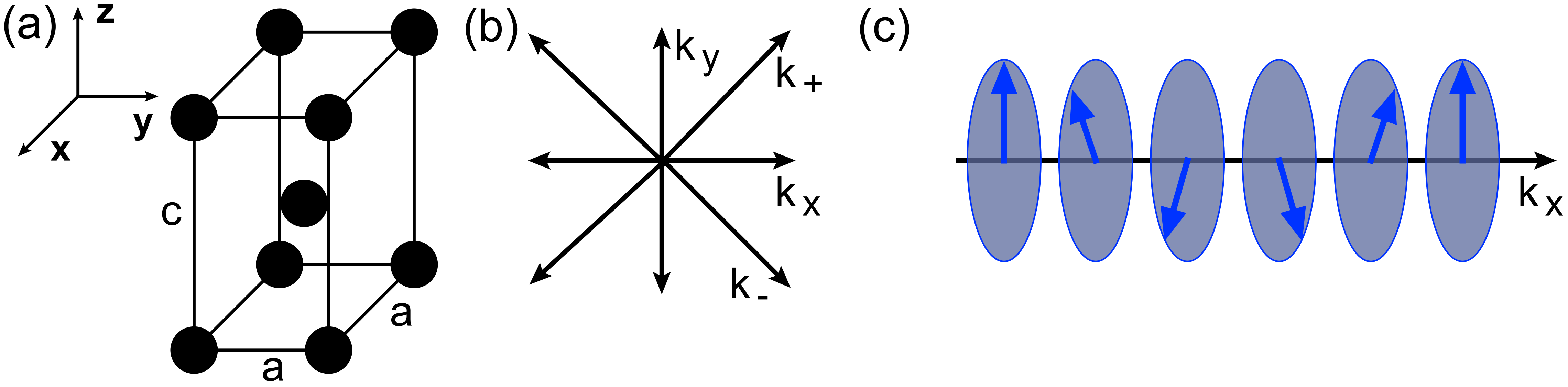}
    \caption{(a) Tetragonal crystalline lattice of \gdru~ where only magnetic Gd$^{3+}$ ions are shown. (b) According to the experimental observations~\cite{khanh2020,khanh2022}, \gdru~ has ordered magnetic phases with modulation vectors $\m{k}_{x,y}$. Here we also show vectors $\m{k}_\pm = \m{k}_x \pm \m{k}_y$. The respective harmonics play an important role in the square skyrmion lattice stabilization. (c) Momentum-dependent biaxial anisotropy (consisting of magnetodipolar, single-ion, and compass anisotropic exchange contributions) favors vertical screw spirals for $\m{k}_{x,y}$ modulations.}
    \label{fig1}
\end{figure*}

In the present study, based on recent experimental findings, we revisit the high-temperature part of the phase diagram of tetragonal frustrated antiferromagnets. We generalize the mean-field approach of Ref.~\cite{utesov2021tetragonal} to include the effects of the biquadratic exchange interaction into consideration. We show that its interplay with complex magnetic anisotropies results in a plethora of phase diagrams in the temperature--magnetic field plane. Using the analytical expressions for various spin structures' free energies, we discuss their competition. It is shown that the existence of higher-order harmonics is crucial for the stabilization of the square SkL due to a peculiar correction to the free energy, which resembles the corresponding term for hexagonal SkLs arising in external field due to the condition $\m{k}_1+\m{k}_2+\m{k}_3=0$ (see, e.g., Refs.~\cite{muhlbauer2009,leonov2015,utesov2022}). In our case, we denote modulation vectors of two principal helical constituents of SkL as $\m{k}_{x,y}$ and the higher harmonics ones as $\m{k}_\pm = \m{k}_x \pm \m{k}_y$. The corresponding condition reads $\m{k}_x \pm \m{k}_y - \m{k}_\pm = 0$. The same conclusion was also emphasized in the recent numerical study~\cite{hayami2025effect}. Exploring the parameter space, we find that in the case of magnetodipolar interaction accompanied by additional easy-axis contribution along the high-symmetry tetragonal axis and in a certain range of biquadratic exchange constant, the phase diagrams similar to the experimental one of~\gdru~\cite{garnier1996giant,khanh2020,khanh2022} can be obtained. Thus, we also obtain a reasonable estimation for the biquadratic exchange strength.
 
The rest of the paper is organized as follows. In Section~\ref{Sform}, we introduce the formalism. Various possible spin structures and relations between their free energies are discussed in Sec.~\ref{Sstruct}. In Sec.~\ref{Sphase}, we present particular phase diagrams, including the one corresponding to~\gdru. Section~\ref{Sdisc} is devoted to a general discussion of the results. Section~\ref{Ssum} contains our conclusions and summary. Finally, two appendices contain detailed information about free energy calculations and topological properties of some spin structures.

\section{General formulae}
\label{Sform}

Following the previous studies~\cite{hayami2021square,utesov2021tetragonal,wang2021}, we consider a model which includes frustrated isotropic exchange, magnetodipolar interaction and/or anisotropic exchange, Zeeman term, and biquadratic exchange. Assuming one magnetic ion in the unit cell, we can write the reciprocal space Hamiltonian as follows:
\begin{eqnarray} \label{ham1}
  \mathcal{H} &=& \mathcal{H}_\textrm{EX} + \mathcal{H}_\textrm{AN} + \mathcal{H}_\textrm{Z} + \mathcal{H}_\textrm{BEX}, \\
  \label{hex}
  \mathcal{H}_\textrm{EX} &=& -\frac12 \sum_\mathbf{q} J_\mathbf{q} \left(\mathbf{S}_\mathbf{q} \cdot \mathbf{S}_{-\mathbf{q}}\right), \\
	\label{hdip}
  \mathcal{H}_\textrm{AN} &=& \frac12 \sum_\mathbf{q} {\cal D}^{\alpha \beta}_\mathbf{q} S^\alpha_\mathbf{q} S^\beta_{-\mathbf{q}}. \\
	\label{hz}
\mathcal{H}_\textrm{Z} &=& - \sqrt{N} \m{h} \cdot \m{S}_\m{0}, \\
	\label{hba}
   \mathcal{H}_\textrm{BEX} &=& \frac{K}{N} \sum_\m{q} (\m{S}_\m{q} \cdot \m{S}_{-\m{q}})^2,
\end{eqnarray}
where $N$ is the number of lattice sites and we measure magnetic fields in energy units, $h = g \mu_\textrm{B} H$. The last term with $K>0$ here is written in the form suggested in Refs.~\cite{hayami2017,hayami2021square} as the most important contribution to the biquadratic exchange in itinerant magnets. The first two terms~\eqref{hex} and~\eqref{hdip} can be combined into a single term \mbox{$\mathcal{H}_\textrm{EX} + \mathcal{H}_\textrm{AN} = -\sum_\m{q} \mathcal{H}^{\alpha \beta}_\m{q} S^\alpha_\m{q} S^\beta_{-\m{q}}$.} Noteworthy, tensor $\mathcal{H}^{\alpha \beta}_\m{q}$ eigenvalues $\lambda_1(\mathbf{q}) \geq \lambda_2(\mathbf{q}) \geq \lambda_3(\mathbf{q})$ and respective eigenvectors $\mathbf{v}_1(\mathbf{q}), \, \mathbf{v}_2(\mathbf{q}), \, \mathbf{v}_3(\mathbf{q})$ define the ``zoo'' of possible spin structures in each particular model.

We aim to discuss experimentally relevant cases, when the possible modulation vectors lie in-plane [we denote them as $\m{k}_x = (k,0,0)$ and $\m{k}_y = (0,k,0)$]. They are equivalent due to tetragonal symmetry~\cite{khanh2020,khanh2022}. These vectors are roughly (up to small anisotropy-induced corrections) determined by the maxima of the exchange interaction Fourier transform. In the discussion below, we will also need $\m{k}_\pm = \m{k}_x \pm \m{k}_y$. In \gdru, $k=0.22 \times 2 \pi/ a$, where $a$ is the in-plane lattice parameter~\cite{khanh2020}. Moreover, the corresponding spin structures have transverse order parameters, and we conclude that the longitudinal directions for $\m{k}_x$ and $\m{k}_y$ are the respective hard axes. Then, the corresponding perpendicular in-plane directions can be easy or middle axes. Importantly, the former case is relevant to magnetodipolar interaction, which should be important for Gd$^{3+}$ ions with zero orbital momentum~\cite{utesov2021tetragonal}. In the general case, the anisotropic part of the Hamiltonian can also include compass anisotropic exchange~\cite{wang2021} and single-ion terms. Fig.~\ref{fig1} illustrates these crucial concepts.

\begin{figure*}
    \centering
    \includegraphics[width=0.3\linewidth]{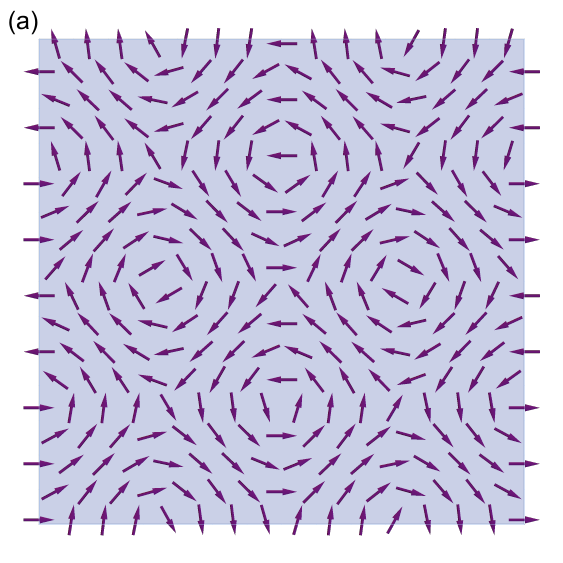}
    \includegraphics[width=0.3\linewidth]{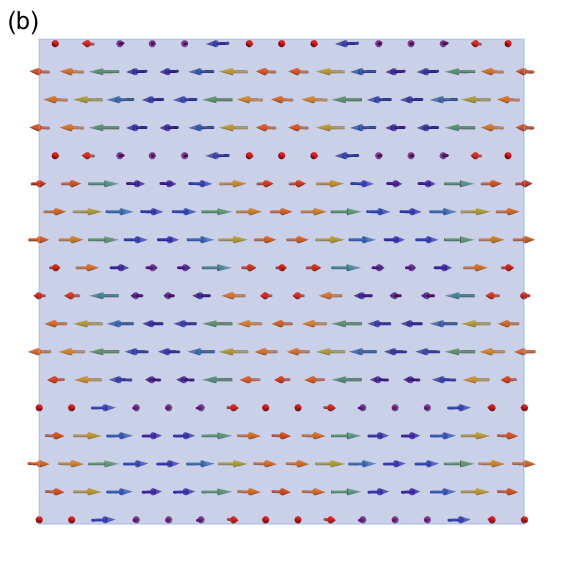}
    \includegraphics[width=0.3\linewidth]{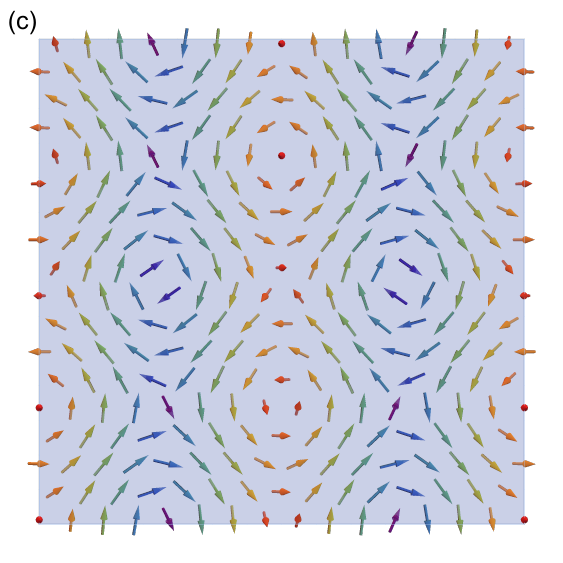}
    \includegraphics[width=0.3\linewidth]{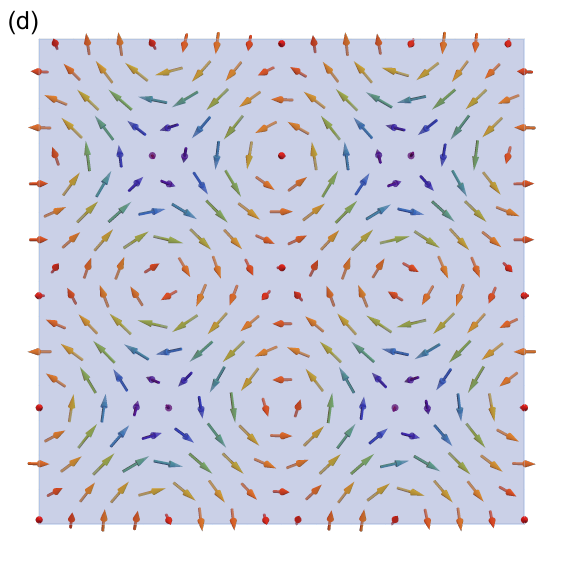}
    \includegraphics[width=0.3\linewidth]{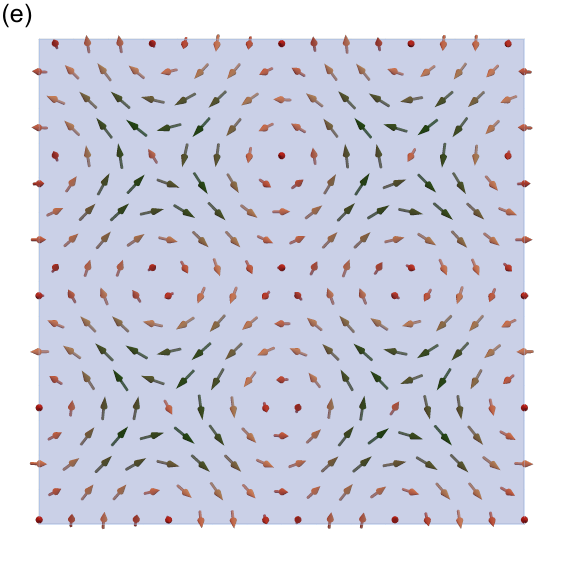}
    \includegraphics[width=0.3\linewidth]{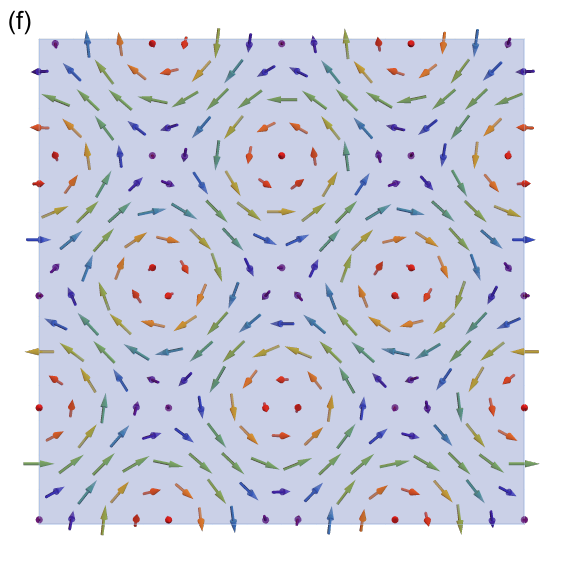}
    \caption{Sketch of main spin structures in the $xy$ plane, which can appear in typical phase diagrams of tetragonal frustrated antiferromagnets. Color encodes $\hat{z}$ projections of spins (red -- ``spin up'', violet -- ``spin down'', green -- spin in the $xy$ plane). (a) Superposition of two spin-density waves with modulation vectors $\m{k}_x$ and $\m{k}_y$ and perpendicular in-plane polarizations (2SXY). In real space, it corresponds to a vortex lattice. (b) Superpositions of two SDWs with in-plane and out-of-plane polarizations (S2YZ). (c) Superposition of screw helicoid with $\m{k}_y$ and SDW with $\m{k}_x$ (1Q+1S); meron -- anti-meron lattice. (d) Topologically-nontrivial square skyrmion lattice made of two screw helicoids with $\m{k}_x$ and $\m{k}_y$, harmonics with $\m{k}_\pm$ polarized along $\hat{z}$, and uniform magnetization (2Q). Its topological charge per unit cell is $n_\textrm{sk}= 1$ (e) Topologically-trivial version of 2Q structure at stronger external field (all $s^z_i>0$), $n_\textrm{sk}=0$. (f) Alternative topologically-nontrivial structure 2Q$^\prime$ with $n_\textrm{sk}=2$.  }
    \label{fig2}
\end{figure*}

For our calculations, we employ the mean-field approach based on the smallness of the thermodynamically average ordered spin value, $s = \langle S \rangle \ll S$. In this case, which corresponds to high temperatures near $T_\textrm{N}$, we can express the free energy as a function of $s_\m{q}$ (reciprocal space) and $\m{s}_i$ (direct space) as follows (see, e.g., Ref.~\cite{gekht1984} for details):
\begin{eqnarray}\label{Free1}
  \mathcal{F} &=& - \sum_\mathbf{q} \mathcal{H}^{\alpha\beta}_\mathbf{q} s^\alpha_\mathbf{q} s^\beta_{-\mathbf{q}} - \sqrt{N} \mathbf{h} \cdot \mathbf{s}_{\bf 0} + A T \sum_i s^2_i \nn \\&& + \frac{K}{N} \sum_\m{q} (\m{s}_\m{q} \cdot \m{s}_{-\m{q}})^2 + B T_\textrm{N} \sum_i s^4_i,
\end{eqnarray}
Expansion parameters $A$ and $B$ are given by
\begin{eqnarray}
  A = \frac{3}{2S(S+1)}, \,
  B = \frac{9[(2S+1)^4-1]}{20 (2S)^4(S+1)^4}.
\end{eqnarray}
For $S=7/2$, one has $A \approx 0.095$ and $B \approx 0.002$. Note that by assuming a certain spin structure, we readily obtain its free energy~\eqref{Free1} in the form of a quartic polynomial suitable for either analytical or numerical minimization~\cite{gekht1984,utesov2021tetragonal,utesov2022}.

\section{Spin structures and phase transitions}
\label{Sstruct}

In the present study, we consider the magnetic field only along the $\m{c}$ axis. Some of the relevant spin structures are shown in Fig.~\ref{fig2}. We distinguish: (i) single-modulated spin-density wave (SDW) which we denote 1S, it turns out to be relevant only for easy axes along $\hat{z}$, (ii) three variants of double-modulated superpositions of two SDWs with different polarizations (2SXY, 2SYZ, 2SZZ), (iii) vertical screw helicoid (1Q), (iv) transverse cone spiral (TC), (v) hybrid structure made of vertical spiral and SDW with perpendicular modulation vector (1Q+1S), (vi) superposition of two vertical spirals (2Q), which is the square skyrmion lattice in a certain range of fields, (vii) alternative 2Q ordering (2Q$^\prime$), where $\hat{z}$ components of the spin structure have $\m{k}_\pm$ modulation vectors. Details of calculations supporting our claims are presented in Appendix~\ref{AFreeEn}.

We start with the easier case of in-plane easy axes for $\m{k}_{x,y}$ modulation vectors. The Néel temperature is given by $T_\textrm{N} = \lambda_1(\m{k}_{x,y})/A$. Within the framework of the mean-field approach, we prove several general statements crucial for the corresponding phase diagrams. The spin-density wave (1S) structure's free energy~\eqref{Af1S} is always larger than the free energy of the superposition of two spin-density waves (vortical phase, 2SXY)~\eqref{Af2Sxy} since
\be
  -\frac{t^2}{2(3 b + K)} > -\frac{t^2}{5 b + K}
\ee
for $K \geq 0$ and $t \equiv A(T_N - T)>0$, where both phases are defined (here $b = B T_\textrm{N}$). So, we can exclude 1S from the analysis. Other variants of 2S structures are also irrelevant due to the axes hierarchy.

Next, we can compare the free energies of the spiral 1Q phase~\eqref{Af1Q} and the hybrid 1Q+1S structure~\eqref{Af1Q1S}, assuming that both of them exist. Remarkably, their difference is always positive:
\begin{widetext}
\be
   f_\textrm{1Q} - f_\textrm{1Q+1S}  = -\frac{4b t^2 - 4 b \Lambda_1 t + (3b+K) \Lambda_1^2}{8b (2 b + K)} + \frac{4b (b + 2K) t^2 - 4 b (b+K)\Lambda_1 t + (b+K)(5b+K) \Lambda_1^2}{8b (2 b^2 + 5 b K+K^2)} =  \\ \nn
   = \frac{(b \Lambda_1 + K t)^2}{2(2b+K)(2 b^2 + 5 b K+K^2)}>0.
\ee
\end{widetext}
Moreover, it can be shown that in the region where 1Q+1S does not exist [$t<(5b+K)\Lambda_1/2b$], but 1Q does [$t>(3b+K)\Lambda_1/2b$], 2SXY has minimal free energy. As a corollary, the 1Q structure does not appear in the phase diagrams.

The same applies if we compare the free energies of 1Q+1S~\eqref{Af1Q1S} and 2Q~\eqref{Af2Q} [without the correction~\eqref{Af2Qcorr} important at moderate magnetic fields]. Note that these phases are defined in the same region $t>(5b+K)\Lambda_1/2b$ and emerge via the second-order phase transition from 2SXY. The difference in the free energies
\be
  f_\textrm{2Q} - f_\textrm{1Q+1S} =  \frac{(2 b t -5 b \Lambda_1 - K \Lambda_1 )^2}{8(9b+2K)(2 b^2 + 5 b K+K^2)} \geq 0. \nn \\
\ee
Equality can be reached only at the phase boundary. We conclude that the correction to the free energy of the 2Q phase~\eqref{Af2Qcorr} is crucial in its stabilization. 

The origin of the correction is the following: in the external field, 2Q develops a homogeneous magnetization component, and the spin structure is as follows:
\be \label{s2Q}
  \m{s}_j &=& s (\hat{y} \sin{\m{k}_x \m{R}_j} + \hat{x} \sin{\m{k}_y \m{R}_j}) \nn \\ &&+\hat{z} [p (\cos{\m{k}_x \m{R}_j} + \cos{\m{k}_y \m{R}_j}) + m], 
\ee
with certain $s, p$ and $m$ (see Appendix~\ref{AFreeEn}). $z$-components of spin structure due to the quartic term in the free energy~\eqref{Free1} lead to instability for the emergence of additional harmonics $\m{k}_\pm$ lowering the total free energy. The physics here resembles the conventional mechanism of the stabilization of hexagonal skyrmion lattices in magnetic field due to the interplay of constant magnetization and three helical constituents with $\m{k}_1+\m{k}_2+\m{k}_3 =0 $ (see, e.g., Refs.~\cite{gekht1984,muhlbauer2009,leonov2015,utesov2022}). In the present case, the crucial terms $\propto m p^2$ emerge due to the equality $\m{k}_x \pm \m{k}_y - \m{k}_\pm = 0$  and similar ones. Notably, the corresponding correction to the spin structure
\be \label{s2Qcorr}
  \delta \m{s}_i = \hat{z} s_\pm \left[ \cos{(\m{k}_+ \m{R}_j + \pi) } + \cos{(\m{k}_- \m{R}_j} + \pi) \right]
\ee
``patches'' the holes (where $\m{s}_j = m \hat{z}$) in the modulated part of~\eqref{s2Q} for $\m{k}_{x,y} \m{R}_j = \pi l_{x,y} $ and $\m{k}_{x} \m{R}_j - \m{k}_{y} \m{R}_j = 2 \pi l + \pi$ (all $l$s are integers). Importance of these harmonics was also highlighted in Refs.~\cite{hayami2022rectangular,hayami2025effect}.

In the external field, the free energy parameters should be altered as follows: $t \rightarrow t^\prime = t - 2 b \chi^2 h^2$ and $\Lambda_1 \rightarrow \Lambda^\prime_1 = \Lambda_1 + 4 b \chi^2 h^2$, where $\chi$ represents the uniform susceptibility. All free energies also acquire a ``paramagnetic'' correction $-\chi h^2/2$. So, in the external field, the above statements are intact. However, note that the correction to the 2Q phase free energy~\eqref{Af2Qcorr} is $\propto m^2$ and becomes crucial at moderate fields for which both $m$ and $p$ are significant.

We conclude that for in-plane easy axes $\Lambda_1 > 0$, the generic phase diagram should consist of (i) 2SXY phase at high $T$ and $h$, (ii) 1Q+1S at low $T$ and $h$, and (iii) 2Q at moderate $h$, as well as paramagnetic phase above $T_\textrm{N}(h)$. The latter is given by
\be
  T_\textrm{N}(h) = \frac{\lambda_1(\m{k}_{x,y}) - 2 b \chi^2 h^2}{A}.
\ee
The curve of second-order phase transitions separating 2SXY and 1Q+1S (or 2Q) is defined as
\be
  t^\prime = \frac{(5 b + K) \Lambda^\prime_1}{2b}.
\ee
The transition between 1Q+1S and 2Q is the first-order one. 

\begin{figure*}
    \centering
    \includegraphics[width=1\linewidth]{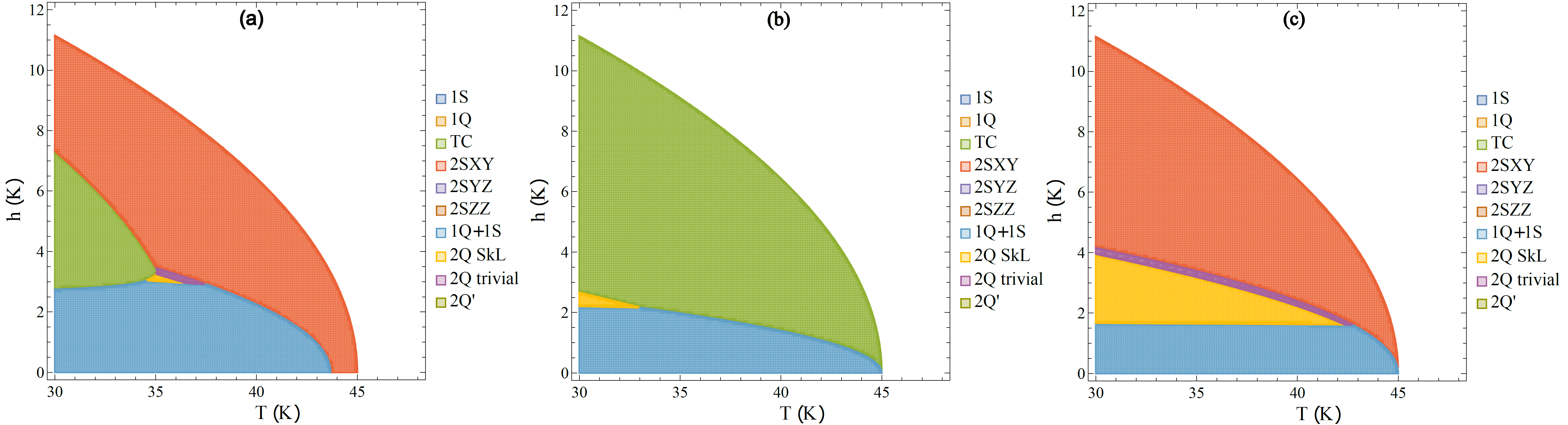}
    \caption{Phase diagrams for simplified models, i.e., for which we include only one of the crucial ingredients. (a) Without the biquadratic exchange, but including anisotropy due to magnetodipolar interaction, the skyrmionic 2Q phase is stable only in a tiny region. At a lower $T$, it loses competition to the transverse conical spiral. (b) For nonzero biquadratic exchange with $K=0.05$~K but without anisotropy, TC is stable in a large region of moderate fields. (c) The same as (b), but $K=0.1$~K. Larger $K$ allows substituting TC with vortical phase 2SXY and makes the square skyrmion lattice stable in the intermediate field region.  }
    \label{fig3}
\end{figure*}

Importantly, near the transition point from 2SXY, for 2Q structure $p \ll 1$ (and, hence, $s_\pm \ll 1$), uniform magnetization $m$ dominates in  $s^z_i$ and 2Q structure is topologically trivial, see Appendix~\ref{ATop} for details. The condition of nonzero topological charge is $2(p+s_\pm) > m$. Note also that the transverse cone spiral, involving the hard magnetization axis, can be stabilized at lower $T$ [it is governed by the $\Lambda_2 > \Lambda_1$ parameter, see Eq.~\eqref{AfTC}]. Its competition with the other phases should be checked separately. Particular examples of phase diagrams are presented below.

When the $\hat{z}$ axis is the easy one, the 1S phase with polarization along $\hat{z}$ can emerge in the high $T$ and low $h$ part of the phase diagram (it competes with the 2SZZ phase there). This can be understood by formally using $\Lambda_1 <0$ in Eqs.~\eqref{Af1Salt} and~\eqref{Af2SZZ}, and noticing that in this case the ordering temperature for 2SYZ and 2SXY is smaller. A magnetic field of a certain value along $\hat{z}$ makes $\Lambda^\prime_1>0$, and the 1S phase can no longer be stable. The same applies to the competition of 1Q and 1Q+1S. According to Eq.~\eqref{Ao1Q1S}, 1Q+1S can only exist if $b \Lambda^\prime_1 + K t^\prime > 0$, which means that for $\Lambda_1 < 0 $ the 1Q phase can appear in the phase diagram. It is also a subject of competition with the 2SYZ phase. As for 1S and 2SZZ, the result depends on the value of the $K/b$ ratio.

\section{Phase diagrams and application to $\text{Gd}\text{Ru}_2\text{Si}_2$}
\label{Sphase}

For the parameter estimation of particular materials, one can use the experimentally measured ordering temperature $T_\textrm{N}$ and the saturation magnetic field of the transition from a noncollinear phase to the field-polarized phase for low temperatures. Neglecting small anisotropic interactions, we have
\be
   J_{\m{k}_{x,y}} \approx 2 T_\textrm{N}/A, \quad h_\textrm{sat} = S (J_{\m{k}_{x,y}} - J_\m{0}).
\ee
These exchange parameters, roughly speaking, determine the overall scales of the axes in the $T-H$ phase diagrams. Particular ``filling'' of the ordered phase region is determined by weak interaction, in our case, biaxial momentum-dependent anisotropy and biquadratic exchange. Without them, our mean-field approach yields a trivial phase diagram consisting of TC and paramagnetic phases. The anisotropy helps us to move TC to low temperatures, and biquadratic exchange further promotes the multiple-modulated structures. We note that for some particular parameter choices, the phase diagrams discussed below correspond well to those obtained numerically in Ref.~\cite{hayami2023bubble}. This further justifies our analytical approach.

Estimations based on \gdru~experimental phase diagram ($T_\textrm{N} \approx 45$~K and $B_\textrm{sat} \approx 10$~T) yield 
\be
  J_{\m{k}_{x,y}} \approx 8.5 \, \textrm{K}, \quad J_\m{0} \approx 4.7 \, \textrm{K}.
\ee
These estimations align with the numerical calculations of Ref.~\cite{bouaziz2022}. Anisotropy of the pure dipolar nature corresponds to~\cite{utesov2021tetragonal}
\be
  \Lambda_1 \approx 0.05 \, \textrm{K}, \quad  \Lambda_2 \approx 0.2\, \textrm{K}.
\ee
Taking also certain value of $J_{\m{k}_\pm}$, for instance, in between $J_{\m{k}_{x,y}}$ and $J_\m{0}$, yields $\Lambda_\pm = 0.95$~K. Using these parameters and $K=0$, we get the phase diagram shown in Fig.~\ref{fig3}(a). In this case, the SkL phase is only stable in a tiny region of the phase diagram. It is instructive to draw also phase diagrams for the isotropic case and nonzero $K$. The results for $K=0.05$~K and $K=0.1$~K are shown in Figs.~\ref{fig3}(b) and (c), respectively. For smaller $K$, the TC structure is stable in the moderate field domain, whereas for larger $K$ it is substituted with 2SXY. Simultaneously, the skyrmion 2Q phase stability domain dramatically expands.

\begin{figure*}
    \centering
    \includegraphics[width=1\linewidth]{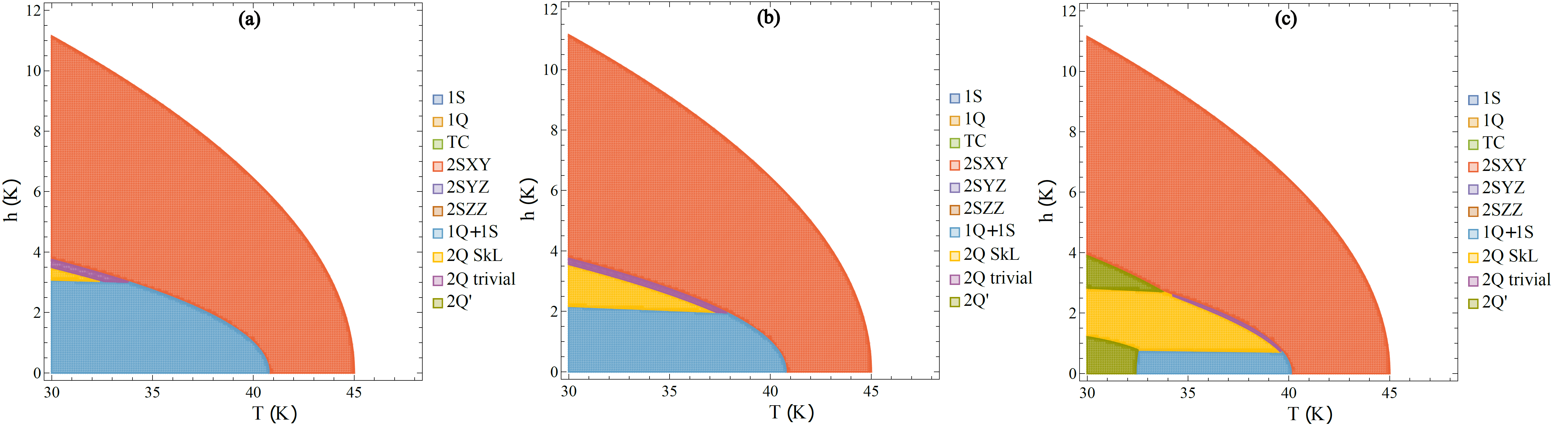}
    \caption{Phase diagrams for the model with both momentum-dependent anisotropy and biquadratic exchange. (a) Even relatively weak biquadratic exchange suppresses single-modulated transverse conical structure TC [cf. Fig.~\ref{fig3}(a)]. (b) Decrease of $\Lambda_\pm$ expands SkL stability domain. (c) Further decrease of $\Lambda_\pm$ and increase of $K$ essentially changes the phase diagram, stabilizing topologically nontrivial 2Q$^\prime$ spin structure with $n_\textrm{sk}=2$.   }
    \label{fig4}
\end{figure*}

We continue playing with the phase diagrams, trying to obtain something similar to the diagram shown in Fig.~1a of Ref.~\cite{khanh2022}. We combine magnetodipolar anisotropy and biquadratic exchange to stabilize 2SXY and suppress TC. We also enhance the anisotropy parameter $\Lambda_1$ to have the temperature of the phase transition between 2S and 1Q+1S about $40$~K using the single-ion easy-plane anisotropy with $Z = 0.1$~K (it makes $\Lambda_1 = 0.15$~K and $\Lambda_2 = 0.2$~K). We arrive at the phase diagram shown in Fig.~\ref{fig4}(a) by taking biquadratic exchange constant $K=0.03$~K, and $J_{\m{k}_\pm}=J_\m{0}$ yielding $\Lambda_\pm = 1.9$~K. Note that the latter value is reasonable according to numerical calculations of Ref.~\cite{bouaziz2022}.
Smaller $\Lambda_\pm = 0.95$ allows expanding the skyrmion lattice stability domain, see Fig.~\ref{fig4}(b). Noteworthy that further decreasing of $\Lambda_\pm$ and increasing of $K$ leads to the emergence of the topologically nontrivial 2Q$^\prime$ phase predicted in Ref.~\cite{wang2021}. This effect is illustrated in Fig.~\ref{fig4}(c) for $\Lambda_\pm=0.5$~K and $K=0.1$~K.

\begin{figure*}
    \centering
    \includegraphics[width=1\linewidth]{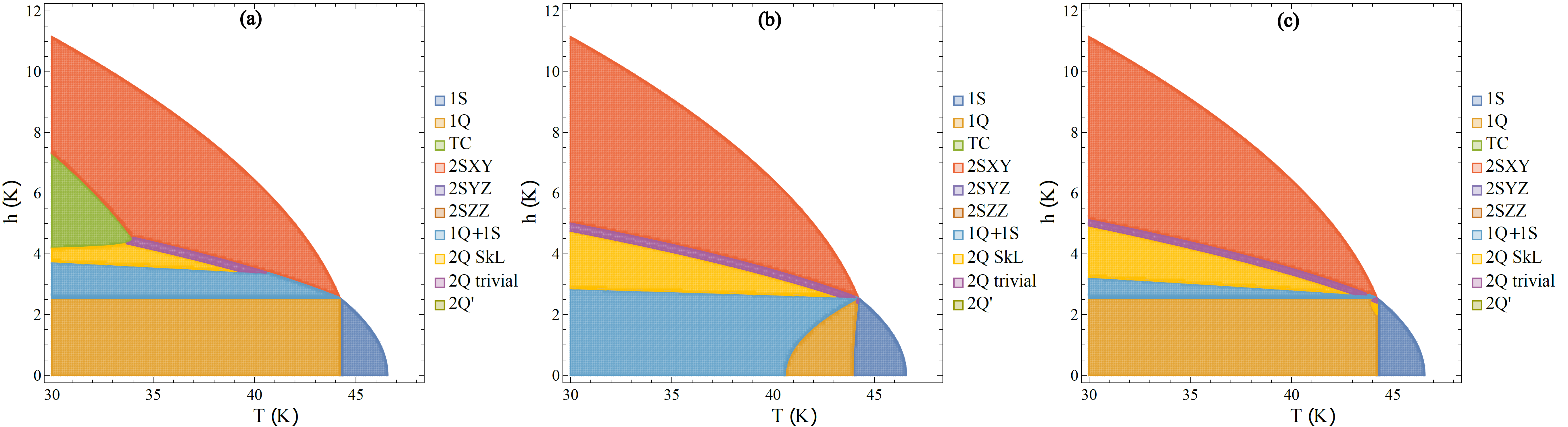}
    \caption{Phase diagrams for the easy axes along the high-symmetry direction $\hat{z}$. (a) Without the biquadratic exchange, helicoid 1Q is stable at weak magnetic fields. (b) Introducing small biquadratic exchange with $K=0.03$~K substantially changes the phase diagram. 
    (c) For $K=0$, enhanced anisotropic exchange and competition among $\m{k}_{x,y}$ and $\m{k}_\pm$ (decreased $\Lambda_\pm$), the well-pronounced skyrmion phase can also be observed.   }
    \label{fig5}
\end{figure*}

The phase diagram shown in Fig.~\ref{fig4}(a) has important similarities with that of Ref.~\cite{khanh2022}. However, there are also several crucial differences. First, we note that according to this study and the early paper~\cite{garnier1996giant}, there should be a phase transition at $H \sim 1$~T between two ordered phases for $T > 40$~K. Second, the phase diagram for $H \parallel [100]$ unambiguously shows that for $h=0$, one of the phases corresponds to the structure we dub 2SYZ. Third, the presaturation phase for $H \parallel [100]$ is SDW with a modulated component polarized along $\hat{z}$. This indicates that the $\hat{z}$ axis is the easy one for $\m{k}_{x,y}$~\cite{utesov2021phase}. Importantly, as discussed above, it leads to essentially different phase diagrams. 

In Fig.~\ref{fig5}(a), we show the result for the dipolar anisotropy with an additional easy-axis term $Z=0.2$~K, $\Lambda_\pm = 0.95$~K, and zero biaxial anisotropy. For such parameters, the single-modulated helicoid 1Q emerges via a first-order phase transition from 1S, and it is stable for weak $h$. A small value of $K$ leads to a shrinkage of the single-modulated phases' stability region and stabilization of 1Q+1S for low temperatures. The result for $K=0.03$~K is shown in Fig.~\ref{fig5}(b); other parameters are the same as above. Finally, by taking $K=0$, $\Lambda_\pm = 0.5$~K, and enhancing the anisotropic exchange ($\Lambda_1 \approx -0.15$~K and $\Lambda_2 \approx 0.4$~K), we arrive at the phase diagram~\ref{fig5}(c) showing expanded SkL stability region even without the biquadratic exchange.

Note that the phase diagram shown in Fig.~\ref{fig5}(c) captures several important features observed for GdRu$_2$Ge$_2$ in Ref.~\cite{yoshimochi2024multistep}. Evidently, in our study, we miss phases with harmonics characterized by additional modulation vectors $(\pm k/2, \pm k/2,0)$. However, even in a simplified calculation, we see that we can discuss phases I, IV, V, and VII using the notations of Ref.~\cite{yoshimochi2024multistep}. Our 1Q+1S structure seems to lose a competition to the hybrid phases II and III. Moreover, in  Ref.~\cite{yoshimochi2024multistep}, the model with $K=0$  and close competition between structures with $\m{k}_{x,y}$ and $(\pm k/2, \pm k/2,0)$ was proposed. According to our findings, the latter is essential for the stabilization of the square SkL in a significant part of the phase diagram.

\begin{figure*}
    \centering
    \includegraphics[width=1\linewidth]{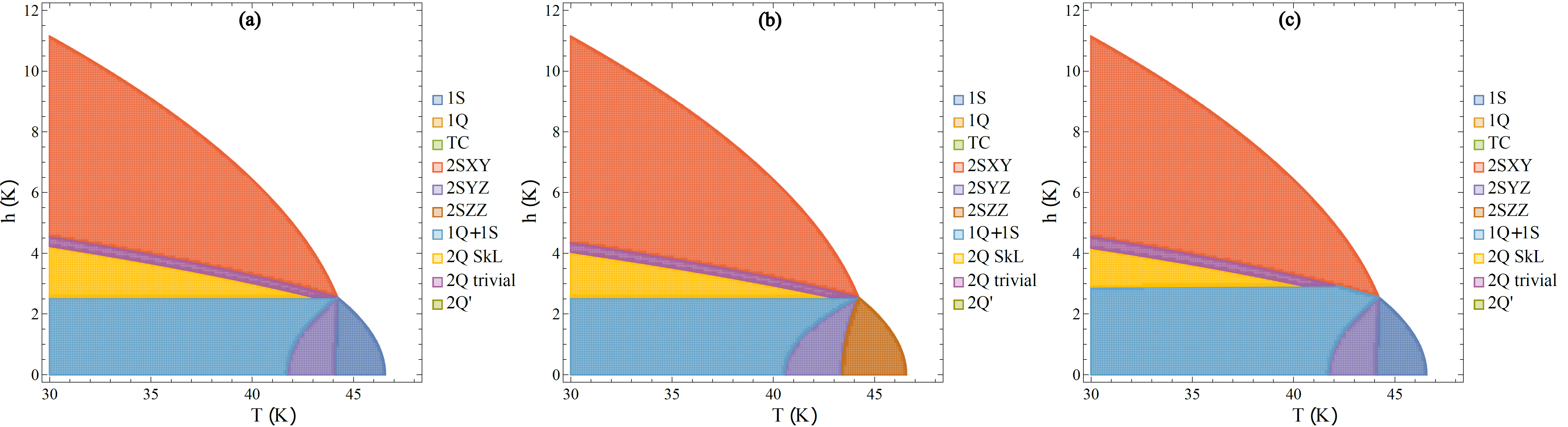}
    \caption{Phase diagrams for the easy axes along the high-symmetry direction $\hat{z}$ and enhanced biquadratic exchange. (a) $K\in(b, 3b)$ leads to stabilization of 2SYZ and 1S at high-$T$ and low-$h$ part of the phase diagram. (b) If $K > 3b$, instead of 1S, the double SDW structure 2SZZ emerges. (c) For $K=0.2$~K, weakened competition among $\m{k}_{x,y}$ and $\m{k}_\pm$ (increased $\Lambda_\pm$), the well-pronounced skyrmion phase can be stabilized at lower temperatures. It is argued that this type of phase diagram corresponds to the real one of~\gdru.   }
    \label{fig6}
\end{figure*}

Now it is evident that to obtain a phase diagram similar to that of~\gdru, we need to increase $K$. As shown in Appendix~\ref{AFreeEn}, if $K>b$ then 1Q loses competition to 2SYZ. This scenario is plausible for the description of the experimental observations. Note that there is another condition: if $K > 3b$ then 1S loses competition to 2SZZ, which, in its turn, also competes with 2SYZ, shrinking its region of stability. Moreover, strong biquadratic exchange leads to stabilization of SkL even at large $\Lambda_\pm$. We illustrate this discussion in Fig.~\ref{fig6}. In panel (a), we take $K=0.2$~K, $\Lambda_\pm = 1.9$~K, and $Z=0.2$~K. In this case, high-$T$ low-$h$ phases are 1S and 2SYZ, and 2Q is stable in the moderate fields region. Taking a larger $K=0.3$~K, we substitute 1S with 2SZZ, see Fig.~\ref{fig6}(b). Finally, to push 2Q to low temperatures and get rid of competition between 2SYZ and 2SZZ, we take $K=0.2$~K, $\Lambda_\pm=5$~K, $Z=0.25$~K and arrive at the phase diagram shown in Fig.~\ref{fig6}(c).

\section{Discussion}
\label{Sdisc}

We believe that the way we obtained the phase diagram shown in Fig.~\ref{fig6}(c) reflects the physics of~\gdru. It requires frustrated exchange interaction $J_\m{q}$ with four maxima at $\pm \m{k}_{x,y}$, much smaller ``diagonal'' exchange $J_{\m{k}_\pm}$, moderate anisotropic exchange with easy axis along $\hat{z}$ and compass in-plane term (in the form of, e.g., magnetodipolar interaction) to stabilize Bloch-type structures, and moderate biquadratic exchange $K$. The latter should be larger than the isotropic Landau coefficient, i.e., $K \gtrsim b = B T_\textrm{N} \sim 0.1$~K. 

There are certain discrepancies between our result Fig.~\ref{fig6}(c) and Fig.~1a of Ref.~\cite{khanh2022}. Here, the thermal fluctuation can play an important role. Using the Ginzburg-Levanyuk criterion~\cite{landau2013statistical}, we obtain
\be
  \textrm{Gi} \sim \frac{B^2}{A^4} \approx 0.04.
\ee
Then, the typical fluctuation region width near second-order phase transition points can be estimated as several Kelvins. That is a possible reason why the 1S phase can be hard to detect experimentally and distinguish from 2SYZ, so the corresponding phase boundary is hidden. Moreover, considering, e.g., the vertical cut of Fig.~\ref{fig6}(c) for $T=43$~K, one can also argue that the intermediate 1Q+1S structure will also be implicit due to strong fluctuations.

Similar arguments apply to the square skyrmion lattice 2Q. We can ``push'' it to lower $T$s by increasing the $\Lambda_\pm$ parameter. However, thermal fluctuations are also a plausible solution, and a narrow wedge of the 2Q SkL phase at $30 \lesssim T <40$~K in Fig.~\ref{fig6}(c) can be hard to distinguish experimentally. Moreover, the topological charge of magnetic unit cells relies on the criterion $2(p+s_\pm)>m$  derived in the continuum spin density approximation. In the lattice picture, if the difference $2(p+s_\pm)- m \equiv \delta s \ll 1$, then for the majority of spins have $s^z_i>0$ and only a few have $s^z_i<0$. On the mean field level, it can be understood by considering the spin structure~\eqref{s2Q} with the correction~\eqref{s2Qcorr}. For $m>0$, the minima of $s^z_i$ correspond to $\m{k}_{x,y} \m{R}_j = \pi + 2 \pi l_{x,y}$. However, due to the incommensurate $\m{k}_{x,y}$, the minimum varies among magnetic unit cells, and only part of them $\propto \delta s$ have at least one spin with $s^z_i <0$. Then, even without fluctuations, in the regions of the phase diagram close to the 2Q trivial phase, the average density of topological charge $\langle n_\textrm{sk} \rangle \ll 1$. In simple words, it means that ``robust'' skyrmions providing a tangible topological Hall effect probably do not exist in the high-$T$ part of the phase diagram, despite the corresponding phase being, in principle, stable.

\section{Summary and conclusion}
\label{Ssum}

Using the high-temperature Landau expansion of the free energy, we analytically discuss possible noncollinear (multiple-)modulated phases of tetragonal frustrated antiferromagnets. We show that the subtle inter-phase competition can be discussed in the framework of a model that includes frustrated exchange coupling, anisotropic exchange, and/or magnetodipolar interaction, single-ion anisotropy, biquadratic exchange, and Zeeman energy. For the exchange coupling, we only assume that it has four equivalent maxima at $\pm \m{k}_{x} = (\pm k,0,0)$ and $\pm \m{k}_{y} = (0,\pm k,0)$ in the reciprocal space. The anisotropies are combined into a momentum-dependent biaxial term. Phase diagram topology essentially depends on the corresponding axes hierarchy.

For the in-plane easy axes case, we proved that a simple helicoid, which is usually a ground state of frustrated systems for weak magnetic fields, is substituted by the structure made of helicoid and spin-density wave with the perpendicular polarization (1Q+1S). Concerning the square skyrmion lattice, we show that it requires an external field and a particular relation among modulation vectors of its main constituents (similar to the standard hexagonal skyrmion lattices). In our case, we need mixed harmonics $\m{k}_x \pm \m{k}_y$ of the basic modulation vectors $\m{k}_{x,y}$. In the developed approach, their amplitude and contribution to the free energy were calculated perturbatively. The resulting phase diagram for generic parameters consists of 1Q+1S, SkL, and vortical phase 2SXY, see Figs.~\ref{fig4}(a) and (b). However, a substantial increase of the biquadratic exchange effective constant $K S^2$ (to tens of percent of the Heisenberg exchange scale) leads to a different phase diagram shown in Fig.~\ref{fig4}(c). In this case, an alternative topologically nontrivial phase emerges.

Easy axes along the high-symmetry direction allow observation of a single-modulated spin-density wave 1S and helicoid 1Q, see Figs.~\ref{fig5}(a), (b), and (c). Notably, 1Q can be stable at low $T$ only for negligible biquadratic exchange. Similarly to the in-plane easy axes case, square SkL is stable at moderate fields. Experimentally, we believe that the case of out-of-plane easy axes sheds some light on the phase diagram of GdRu$_2$Ge$_2$~\cite{yoshimochi2024multistep}. However, the observed phase diagram in that paper is much more complicated due to a close competition between two sorts of modulation vectors, and requires a separate study. 

For enhanced biquadratic exchange, we observe stabilization of SDW superpositions 2SYZ and 2SZZ at high $T$, see Figs.~\ref{fig6}(a), (b), and (c). Notably, it can be argued that 2SYZ should emerge in~\gdru~phase diagram (see Refs.~\cite{garnier1996giant,khanh2022}). Then, by choosing reasonable parameters, we obtain the phase diagram shown in Fig.~\ref{fig6}(c), which seems to reproduce all crucial observations for~\gdru~\cite{khanh2020,khanh2022}.

To conclude, by generalizing previous theoretical studies, we adapt a simple yet fruitful mean-field approach for the high-temperature part of the phase diagrams of tetragonal frustrated antiferromagnets. Interplay of complex magnetic anisotropies and biquadratic exchange leads to a variety of possible phase diagrams. Despite the square skyrmion lattice being stable even if only one of the ingredients is present, their combined effect allows for promoting it to a larger region of the phase diagram. By comparing our findings with experimental phase diagrams of relevant materials~\gdru~ and GdRu$_2$Ge$_2$, we can draw important conclusions about the corresponding microscopic parameters. In particular, for the former we predict the relative strength of the biquadratic exchange of the order of tens of percent of the corresponding Heisenberg exchange, whereas for the latter the biquadratic exchange should be negligible in comparison with the anisotropic terms, which parallels numerics of Ref.~\cite{yoshimochi2024multistep}.

%\section*{CRediT authorship contribution statement}
%
%{\bf Oleg I. Utesov:} Conceptualization, Investigation, Methodology, Visualization,
%Writing – original draft, Writing – review \& editing. {\bf Danila P. Budylev:} Investigation, Visualization.

%\section*{Declaration of competing interest}

%The authors declare that they have no known competing financial interests or personal relationships that could have appeared to influence the work reported in this paper.

\section*{Data availability}

Supporting data are available from the corresponding author upon reasonable request.

\section*{Acknowledgements}

We are grateful to T. Ziman for valuable discussions. O. I. U. acknowledges financial support from the Institute for Basic Science (IBS) in the Republic of Korea through Project No. IBS-R024-D1.

\appendix

\begin{widetext}

\section{Free energies of spin structures}
\label{AFreeEn}

Here, we discuss the details of the analytical calculations and present the corresponding results. The proposed phases are mostly consist of spin-density waves (S) and screw helicoids (Q) with modulation vectors $\m{k}_x$ and $\m{k}_y$. This way, we can discuss all the relevant phases proposed and observed in numerical calculations~\cite{hayami2021square,wang2021,hayami2022rectangular,hayami2023bubble}, as well as distinguished experimentally~\cite{khanh2020, khanh2022}. Importantly, we consider only incommensurate spin structures, allowing us to freely choose the phases of the corresponding constituents (up to certain limitations; see discussion of the 2Q structure below). Magnetic field is oriented along $\hat{z}$ and its main effect is the homogeneous magnetization component $m$.

\subsection{Single-modulated spin-density wave (1S)}

For the in-plane easy axis and modulation vector $\m{k}_x$:
\be \label{As1S}
  \m{s}_j = s \hat{y} \cos{\m{k}_x \m{R}_j} + m \hat{z}.
\ee
We plug this ansatz into Eq.~\eqref{Free1} and obtain
\be
  f = - h m + (A T - \lambda_0) m^2 -\frac{t}{2}s^2  + \frac{K}{8}s^4  + \frac{b}{8}\left( 8 m^4 + 8 m^2 s^2 + 3 s^4 \right),
\ee
where $f =\mathcal{F}/N$, $t = \lambda_1 - A T$, $b = B T_\textrm{N}$, $\lambda_1 \equiv \lambda_1(\m{k}_{x,y})$ is the largest eigenvalue of $\mathcal{H}^{\alpha\beta}_\m{q}$ among all $\m{q}$, and $\lambda_0$ corresponds to $z$ eigenvalue for $\m{q}=0$. The former can also include the demagnetization factor, i.e., $\lambda_0 = (J_\m{0} - \om_0 \mathcal{N}_{zz})/2$ with $\mathcal{N}_{zz}$ being the corresponding demagnetization factor, and $\om_0 = 4 \pi (g \mu_B)^2/v_0$ ($v_0$ being the unit cell volume). Evidently, for $h=0$, we have the ordering temperature $T_\textrm{N} = \lambda_1/A$.

For analytical treatment, assuming that the system is far from the instability towards the ferromagnetic state, we can use the following approximation:
\be
  m = \chi h, \quad \chi \equiv \frac{1}{2(\lambda_1 - \lambda_0)}.
\ee
Then, the magnetic field changes the effective temperature for the order parameter $s$:
\be
  t^\prime = t - 2 b \chi^2 h^2.
\ee
Finally, the minimization yields
\be
  s^2 = \frac{2 t^\prime}{3 b + K}
\ee
and
\be \label{Af1S}
  f_\textrm{1S} = -\frac{(t^\prime)^2}{2(3 b + K)} - \frac{\chi h^2}{2}, \, t^\prime >0.
\ee
Notably, 1S with in-plane polarization always loses the competition to another spin structure, 2SXY, presented below.

We can also consider an alternative version of the 1S structure with polarization along $\hat{z}$ axis:
\be \label{As1Salt}
  \m{s}_j = \hat{z} (s \cos{\m{k}_x \m{R}_j} + m).
\ee
It is a relevant phase for the case of the easy axis along the $\hat{z}$ axis. To formally use the results above, we can think about this structure in the case of the in-plane easy axis. Then, using $\Lambda_1 \equiv \lambda_1(\m{k}_x) - \lambda_2(\m{k}_x)$ (difference between anisotropies for in-plane easy axis and out-of-plane middle axis) we write
\be 
  s^2 = \frac{2 (t^\prime-\Lambda^\prime_1)}{3 b + K}
\ee
and
\be \label{Af1Salt}
  f_\textrm{1S} = -\frac{(t^\prime-\Lambda^\prime_1)^2}{2(3 b + K)} - \frac{\chi h^2}{2}, \, t^\prime >\Lambda^\prime_1,
\ee
where we used magnetic field-dependent $\Lambda^\prime_1 = \Lambda_1 + 4 b \chi^2 h^2$. Comparing this result with~\eqref{Af1S}, we conclude that the structure~\eqref{As1Salt} can emerge only for $\Lambda_1<0$ for magnetic fields such that $\Lambda^\prime_1 <0$. 

\subsection{Double-modulated superposition of spin-density waves (2S)}

For in-plane easy axes, it is represented by the vortical spin structure defined as
\be \label{As2S}
  \m{s}_j &=& s (\hat{y} \sin{\m{k}_x \m{R}_j} + \hat{x} \sin{\m{k}_y \m{R}_j})/\sqrt{2} + \hat{z} m. 
\ee
The function to be minimized reads
\be
  f = - h m + (A T - \lambda_0) m^2 -\frac{t}{2}s^2+ \frac{K}{16}s^4 + \frac{b}{16}\left( 16 m^4 + 16 m^2 s^2 + 5 s^4 \right).
\ee
Using the same tricks as for the 1S structure, we have
\be
  s^2 = \frac{4 t^\prime}{5 b + K}
\ee
and
\be \label{Af2Sxy}
  f_\textrm{2SXY} = -\frac{(t^\prime)^2}{5 b + K} - \frac{\chi h^2}{2}, \, t^\prime>0.
\ee

For the easy-axis along $\hat{z}$, there are two other possibilities. Both spin-density waves can be polarized along the easy axis, 
\be \label{As2Szz}
  \m{s}_j &=& s \hat{z} ( \sin{\m{k}_x \m{R}_j} +  \sin{\m{k}_y \m{R}_j})/\sqrt{2} + \hat{z} m. 
\ee
This spin structure can be referred to as a ``magnetic bubble crystal''. It was addressed in detail in Ref.~\cite{hayami2023bubble}. In this case,
\be
  s^2 = \frac{2 (t^\prime-\Lambda^\prime_1)}{9 b + K}
\ee
and
\be \label{Af2SZZ}
  f_\textrm{2SZZ} = -\frac{(t^\prime-\Lambda^\prime_1)^2}{9 b + K} - \frac{\chi h^2}{2}, \, t^\prime>\Lambda^\prime_1.
\ee
Comparison with free energy~\eqref{Af1Salt} shows that this structure can only be stable for large enough $K> 3b$.

Another possibility is
\be \label{As2SYZ}
  \m{s}_j &=& s \hat{y} \sin{\m{k}_x \m{R}_j} + \hat{z}( p \sin{\m{k}_y \m{R}_j} + m), 
\ee
where one SDW is polarized along $\hat{z}$ and another is in-plane. Its free energy reads
\be 
  f = - h m + (A T - \lambda_0) m^2 -\frac{t}{2}s^2 -\frac{t-\Lambda_1}{2}p^2 + \frac{b}{8}\left[ 8 m^4 + 8 m^2 (s^2 + 3 p^2) + 3 s^4 + 4 s^2 p^2 + 3 p^4\right] + \frac{K}{8}(s^4+p^4). \nonumber \\
\ee
Minimization yields
\be
  s^2 = \frac{2[2 b \Lambda^\prime_1 + (b+K) t^\prime]}{( b+K)(5 b +K)}, \, p^2 = \frac{2[ (b+K) t^\prime -  \Lambda^\prime_1(3b+K)]}{( b+K)(5 b +K)}
\ee
and
\be \label{Af2SYZ}
  f_\textrm{2SYZ} = -\frac{2(b+K) (t^\prime)^2 - 2( b+K) \Lambda^\prime_1 t^\prime + (3b+K) (\Lambda^\prime_1)^2}{2(b+K) (5 b + K)} - \frac{\chi h^2}{2}, \, t^\prime  > - \frac{2 b \Lambda^\prime_1}{b+K} \quad \textrm{and} \quad t^\prime>\frac{\Lambda^\prime_1 (3 b+K)}{b+K}.
\ee

\subsection{Vertical spiral (1Q)}

Here, spins rotate in the plane perpendicular to the modulation vector. The spin structure is defined as
\be \label{As1Q}
  \m{s}_j &=& s \hat{y} \sin{\m{k}_x \m{R}_j} + \hat{z} (m + p \cos{\m{k}_x \m{R}_j}), 
\ee
and its free energy reads
\be 
  f = - h m + (A T - \lambda_0) m^2 -\frac{t}{2}s^2 -\frac{t-\Lambda_1}{2}p^2 + \frac{b}{8}\left[ 8 m^4 + 8 m^2 (s^2 + 3 p^2) + 3 s^4 +2 s^2 p^2 + 3 p^4\right] + \frac{K}{8}(s^2+p^2)^2. \nonumber \\
\ee
In the magnetic field, we use the same $t^\prime$ and  $\Lambda^\prime_1$ as before. 

Minimization yields
\be
  s^2 = \frac{2 b t^\prime + \Lambda^\prime_1 (b+K)}{2 b (2 b+K)}, \, p^2 = \frac{2 b t^\prime  - \Lambda^\prime_1 (3 b+K)}{2 b (2 b+K)}
\ee
and
\be \label{Af1Q}
  f_\textrm{1Q} = -\frac{4b (t^\prime)^2 - 4 b \Lambda^\prime_1 t^\prime + (3b+K) (\Lambda^\prime_1)^2}{8b (2 b + K)} - \frac{\chi h^2}{2}, \, t^\prime  > - \Lambda^\prime_1 (b+K)/2b \quad \textrm{and} \quad t^\prime>\Lambda^\prime_1 (3 b+K)/2b.
\ee
Apparently, the 1Q structure can be stable only if $t^\prime$ is larger than a certain value. Moreover, direct comparison of the free energies with 2SYZ [see Eq.~\eqref{Af2Sxy}] shows that the latter is favorable if $K>b$.

\subsection{Transverse conical spiral (TC)}

In this single-modulated structure, spins rotate in the $xy$ plane:
\be \label{AsTC}
  \m{s}_j &=& s \hat{y} \sin{\m{k}_x \m{R}_j} + p \hat{x} \cos{\m{k}_x \m{R}_j} +  \hat{z} m. 
\ee
The free energy includes another parameter, $\Lambda_2 \equiv \lambda_1(\m{k}_x) - \lambda_3(\m{k}_x)$:
\be 
  f = - h m + (A T - \lambda_0) m^2 -\frac{t}{2}s^2 -\frac{t-\Lambda_2}{2}p^2 + \frac{K}{8}(s^2+p^2)^2 + \frac{b}{8}\left[ 8 m^4 + 8 m^2 (s^2 +  p^2) + 3 s^4 +2 s^2 p^2 + 3 p^4\right] . \nonumber \\
\ee
In contrast to 1Q, parameter $\Lambda_2$ stays intact upon magnetic field variation. The result of minimization reads
\be
  s^2 = \frac{2 b t^\prime + \Lambda_2 (b+K)}{2 b (2 b+K)}, \, p^2 = \frac{2 b t^\prime  - \Lambda_2 (3 b+K)}{2 b (2 b+K)}
\ee
and
\be \label{AfTC}
  f_\textrm{TC} = -\frac{4b (t^\prime)^2 - 4 b \Lambda_2 t + (3b+K) (\Lambda_2)^2}{8b (2 b + K)} - \frac{\chi h^2}{2}, \, t^\prime > - \Lambda_2 (b+K)/2b \quad \textrm{and} \quad t^\prime>\Lambda_2 (3 b+K)/2b.
\ee

\subsection{Double-modulated superposition of helicoid and spin-density wave (1Q+1S)}

This structure consists of a vertical spiral and spin-density wave with the perpendicular modulation vector:
\be \label{As1Q1S}
  \m{s}_j = s_x \hat{y} \sin{\m{k}_x \m{R}_j} + \hat{z} (p \cos{\m{k}_x \m{R}_j} + m) + s_y \hat{x} \sin{\m{k}_y \m{R}_j}. 
\ee
After some calculations, we obtain
\be \nonumber
  f &=& - h m  + (A T - \lambda_0) m^2 -\frac{t}{2}(s^2_x+s^2_y) -\frac{t-\Lambda_1}{2}p^2 +  \frac{K}{8}\left[(s^2_x+p^2)^2 +s^4_y \right] \\ &+& \frac{b}{8}\left[ 8 m^4 + 8 m^2 (s^2_x + 3 p^2 + s^2_y) + 3 s^4_x +2 s^2_x p^2 + 3 p^4 + 3s^4_y + 4 s^2_y (s^2_x + p^2)\right].
\ee
Minimization yields
\be \label{Ao1Q1S}
  s^2_x = \frac{2 b (b + K) t^\prime + \Lambda^\prime_1 (-b^2+ 4 b K+K^2)}{2 b (2 b^2 + 5 b K+K^2)}, \, p^2 = \frac{[2 b t^\prime - \Lambda^\prime_1 (5 b+K)](b + K)}{2 b (2 b^2 + 5 b K+K^2)}, 
 s^2_y = \frac{2 (b \Lambda^\prime_1 + K t^\prime) }{2 b^2 + 5 b K+K^2}
\ee
and
\be \label{Af1Q1S}
  f_\textrm{1Q+1S} = -\frac{4b (b + 2K) (t^\prime)^2 - 4 b (b+K)\Lambda^\prime_1 t^\prime + (b+K)(5b+K) (\Lambda^\prime_1)^2}{8b (2 b^2 + 5 b K+K^2)} - \frac{\chi h^2}{2}.
\ee
All squared order parameters should be positive.

\subsection{Square skyrmion lattice (2Q)}

This spin structure consists of two vertical spirals with perpendicular modulation vectors and a constant magnetization component:
\be \label{As2Q}
  \m{s}_j &=& s (\hat{y} \sin{\m{k}_x \m{R}_j} + \hat{x} \sin{\m{k}_y \m{R}_j})/\sqrt{2} +\hat{z} [p (\cos{\m{k}_x \m{R}_j} + \cos{\m{k}_y \m{R}_j})/\sqrt{2} + m]. 
\ee
For $m \neq 0$, it represents a topologically nontrivial square skyrmion lattice. The corresponding free energy function reads
\be \nonumber
  f= - h m + (A T - \lambda_0) m^2 -\frac{t}{2}s^2 -\frac{t-\Lambda_1}{2}p^2 + \frac{b}{16}\left[ 16 m^4 + 16 m^2 (s^2 + 3 p^2) + 5 s^4 +6 s^2 p^2 + 9 p^4\right] + \frac{K}{16}(s^2+p^2)^2. \\
\ee
Minimization yields
\be
  s^2 = \frac{6 b t^\prime + \Lambda^\prime_1 (3 b+K)}{b (9 b+2 K)}, \, p^2 = \frac{2 b t^\prime - \Lambda^\prime_1 (5 b+K)}{b (9 b+2 K)}
\ee
and
\be \label{Af2Q}
  f_\textrm{2Q} = -\frac{8b (t^\prime)^2 - 4 b \Lambda^\prime_1 t^\prime + (5b+K) (\Lambda^\prime_1)^2}{4b (9 b + 2 K)} - \frac{\chi h^2}{2}, \, t^\prime > - \Lambda^\prime_1 (3b+K)/6b,  \quad \textrm{and} \quad t^\prime >\Lambda^\prime_1 (5 b+K)/2b.
\ee

However, in contrast to other spin structures above, this structure possesses a crucial instability towards the formation of additional harmonics with $\m{k}_\pm$ due to the quartic nonlinearity. To get an insight into this phenomenon, we add the following correction to the spin structure~\eqref{As2Q}:
\be \label{As2Qcorr}
  \hat{z} \left[ s_+ \cos{(\m{k}_+ \m{R}_j + \varphi_+) } + s_- \cos{(\m{k}_- \m{R}_j} +\varphi_-) \right].
\ee
Indeed, it is easy to see that $z$-components of the order parameter~\eqref{As2Q} produce linear in $s_\pm$ terms in free energy for $m \neq 0$
\be
  3 b m p^2 (s_+ \cos{\varphi_+} + s_- \cos{\varphi_-})
\ee
Assuming $m>0$ we see, that the choice $\varphi_\pm = \pi$ results in some positive nonzero $s_\pm = q/\sqrt{2}$ (these modes repulsion is negligible). With these additional harmonics, the analytical minimization of the free energy is cumbersome, and instead, we use a perturbative treatment. Based on the obtained values for $s$ and $p$, we expand the free energy in series in $q$ up to $O(q^2)$. Then, the value of $q$ can be easily found. We obtained
\be
  q = \frac{3 \sqrt{2} m (5 b \Lambda^\prime_1 + K \Lambda^\prime_1 - 2 b t^\prime)}{3 b (7 \Lambda^\prime_1 - 3\Lambda^\prime_\pm - t^\prime) + 2 K (2 \Lambda^\prime_1 - \Lambda^\prime_\pm + t^\prime) },
\ee
where the parameter $\Lambda^\prime_\pm = \lambda_1(\m{k}_x) - \lambda_z(\m{k}_\pm) + 4 b \chi^2 h^2$ is introduced. In the general case, since $\m{k}_\pm$ do not correspond to the maxima of the exchange interaction Fourier transform, one can expect that $\Lambda_\pm \gg \Lambda_{1,2}$. For the 2Q spin structure free energy correction, we get
\be \label{Af2Qcorr}
  \delta f_\textrm{2Q} = - \frac{9 m^2 (2 b t^\prime -5 b \Lambda^\prime_1 - K \Lambda^\prime_1 )^2}{(9b + 2K)[t^\prime(3b-2K) - \Lambda^\prime_1 (21b+4K) + \Lambda^\prime_\pm (9b +2 K) ]}.
\ee

\subsection{Alternative 2Q ordering (2Q$^\prime$)}

The possibility of this spin structure emergence was predicted in Ref.~\cite{wang2021}. In real space, it is given by
\be \label{As2Qalt}
  \m{s}_j = s (\hat{y} \sin{\m{k}_x \m{R}_j} + \hat{x} \sin{\m{k}_y \m{R}_j})/\sqrt{2} + \hat{z} [ p (\cos{\m{k}_+ \m{R}_j} + \cos{\m{k}_- \m{R}_j})/\sqrt{2} + m], \nonumber
\ee
with the free energy
\be \nonumber
  f &=& - h m  + (A T - \lambda_0) m^2 -\frac{t}{2}s^2 -\frac{t-\Lambda_2}{2}p^2 +  \frac{K}{16}(s^4+p^4) + \frac{b}{16}\left[ 16 m^4 + 16 m^2 (s^2 + 3 p^2) + 5 s^4 + 4 s^2 p^2 + 9 p^4\right]. \\
\ee
We note that the simultaneous existence of both $\m{k}_\pm$ harmonics allows obtaining the minimal free energy here due to negative contributions from terms, for instance, including $\m{k}_x,\m{k}_x,-\m{k}_+,-\m{k}_-$. Minimization yields
\be
  s^2 = 4\frac{7 b t^\prime + K t^\prime + 2 \Lambda^\prime_2 b}{41b^2+14bK+K^2}, \, p^2 = 4 \frac{3 b t^\prime + K t^\prime - \Lambda_2 (5 b+K)}{41b^2+14bK+K^2}
\ee
and
\be \label{Af2Qalt}
  f_{\textrm{2Q}^\prime} = -\frac{2 (5 b+K) (t^\prime)^2 - 2 (3b +K)\Lambda_2 t^\prime + (5b+K) (\Lambda^\prime_2)^2}{41b^2+14bK+K^2} - \frac{\chi h^2}{2}, \, t^\prime > - 
 \frac{2 \Lambda^\prime_2 b}{7b+K}, \quad \textrm{and} \quad  t^\prime > \frac{\Lambda^\prime_2 (5 b+K)}{3b+K}.
\ee

\section{Topological properties of multiple-modulated spin structures}
\label{ATop}

Here we briefly discuss the topological properties of the considered double-modulated spin structures, in particular 1Q+1S, 2Q, and 2Q$^\prime$. Instead of the total topological charge $Q$~\eqref{charge1}, we consider its density
\be
  \rho = \frac{1}{4 \pi} \m{n} \cdot [\partial_x \m{n} \times \partial_y \m{n}]
\ee
and it's integral over the magnetic unit cell, $n_\textrm{sk}$. Here $\m{n}$ is normalized to unity spin $\m{s}$. Note that we have an underlying lattice structure, so we need to interpolate the discrete set $\m{s}_i$ to  $\m{s}(x,y)$. Tedious calculations yield
\be
  \rho_\textrm{1Q+1S}(\m{r}) = - \frac{s_x s_y \cos{\m{k}_y \m{r}}(p+ m \cos{\m{k}_x \m{r}})}{\left[ s^2_x \sin^2{\m{k}_x \m{r}} + s^2_y \sin^2{\m{k}_y \m{r}} + (p \cos{\m{k}_x \m{r}} + m )^2\right]^{3/2}}
\ee
for spin structure~\eqref{As1Q1S}, and
\be
 \rho_\textrm{2Q}(\m{r}) = -\frac{2 s^2 \left[ \sqrt{2} p \left(\cos{\m{k}_x \m{r}} + \cos{\m{k}_y \m{r}} \right)  + 2 m  \cos{\m{k}_x \m{r}} \cos{\m{k}_y \m{r}} - q \frac{ 6 + 2\cos{2\m{k}_x \m{r}} +2\cos{2\m{k}_y \m{r}} - \cos{2\left(\m{k}_x + \m{k}_y \right)\m{r} }- \cos{2\left(\m{k}_x - \m{k}_y \right)\m{r} }}{2 \sqrt{2}}\right]}{\left[  2 s^2(\sin^2{\m{k}_x \m{r}} + \sin^2{\m{k}_y  \m{r}}) + \left\{2 m + \sqrt{2} p ( \cos{\m{k}_x \m{r}} + \cos{\m{k}_y \m{r}}) -2 \sqrt{2} q \cos{\m{k}_x \m{r}} \cos{\m{k}_y \m{r}}\right\}^2 \right]^{3/2}} \nn \\
\ee
for spin structure~\eqref{As2Q} with correction~\eqref{As2Qcorr}. These equations can also be utilized for 2S and 2Q$^\prime$ phases with a proper choice of the order parameters.

We plot densities of topological charge for 1Q+1S, 2Q, and 2Q$^\prime$ for certain reasonable parameters in Fig.~\ref{figA1} for a domain containing $2 \times 2$ magnetic unit cells. 1Q+1S structure has zero net topological charge. However, there are peculiar stripes of topological charge due to meron-antimeron spin structure~\cite{khanh2022,wood2023}. Square skyrmion lattice 2Q has $n_\textrm{sk}=\pm 1$, whereas the alternative structure 2Q$^\prime$ has $n_\textrm{sk}=\pm 2$~\cite{hayami2021square,wang2021}.

\begin{figure*}
    \centering
    \includegraphics[width=0.29\linewidth]{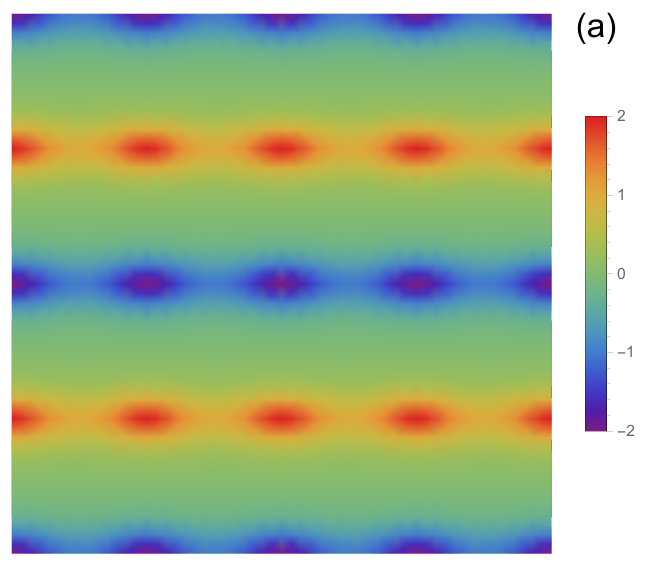}
    \includegraphics[width=0.29\linewidth]{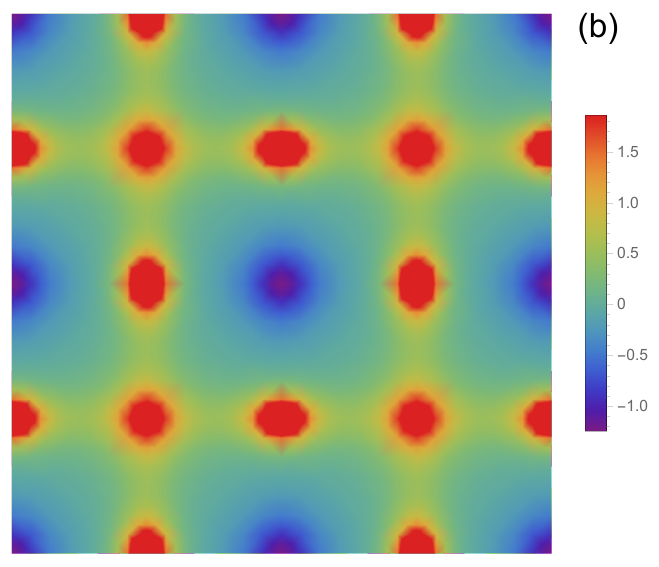}
    \includegraphics[width=0.29\linewidth]{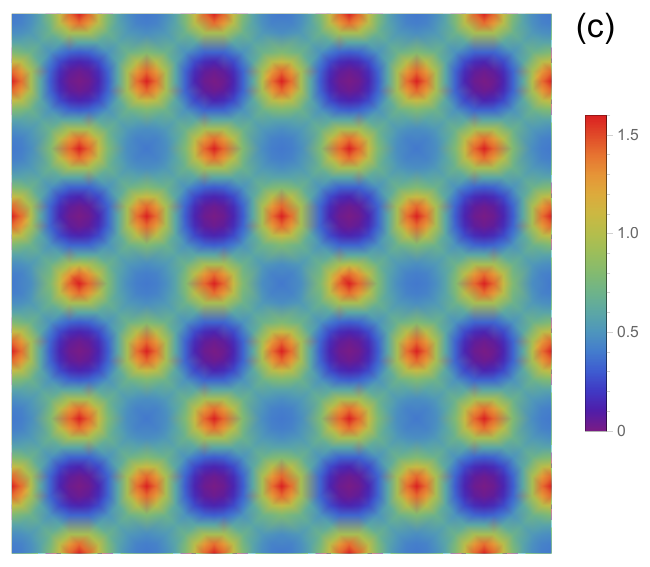}
    \caption{Density of topological charge for double-modulated spin structures in $2 \times 2$ magnetic unit cells domain of direct space. (a) 1Q+1S structure, (b) square skyrmion lattice 2Q stabilized at moderate magnetic fields, and (c) alternative double-modulated 2Q$^\prime$ structure which can be stable for strong biquadratic exchange even at zero magnetic field.  }
    \label{figA1}
\end{figure*}

\end{widetext}

\bibliography{TAFbib}

\end{document}